# Decorrelation of internal quantum efficiency and lasing threshold in AlGaN-based separate confinement heterostructures for UV emission

Sergi Cuesta,[1,a)] Lou Denaix,[1,a)] Le Si Dang,[2] and Eva Monroy[1]

[1] Univ. Grenoble-Alpes, CEA, Grenoble INP, IRIG, PHELIQS, 17-avenue des Martyrs, 38000 Grenoble, France

[2] Univ. Grenoble-Alpes, CNRS, Institut Néel, 25-avenue des Martyrs, 38000 Grenoble, France

**ABSTRACT**

In this paper, we study the internal quantum efficiency and lasing threshold of AlGaN/GaN separate confinement heterostructures designed for ultraviolet laser emission. We discuss the effect of carrier localization and carrier diffusion on the optical performance. The implementation of graded index separate confinement heterostructures results in an improved carrier collection at the multi-quantum well, which facilitates population inversion and reduces the lasing threshold. However, this improvement is not correlated with the internal quantum efficiency of the spontaneous emission. We show that carrier localization at alloy inhomogeneities results in an enhancement of the radiative efficiency but does not reduce the laser threshold, more sensitive to the carrier injection efficiency.

---

a) S. Cuesta and L. Denaix contributed equally to this work.

AlGaN is a promising material to for the fabrication of ultraviolet (UV) lasers in the range of 200-365 nm.[1] However, the implementation of AlGaN laser diodes has encountered technical challenges to achieve efficient current injection and p-type doping, which lead to high threshold devices. An alternative approach is to pump directly the semiconductor structure with a pulsed laser[2–4] or a high energy electron beam[5,6], obviating the need of doping or contacts. In this context, there is a strong interest in designing new architectures with reduced lasing threshold. Thinking of an edge emitting laser consisting of a separate confinement heterostructure (SCH), the quantum wells and the waveguide must be treated as a whole. The waveguide must be designed to provide a high optical confinement factor in the wells, and its structural quality is important to minimize the internal losses due to light scattering. Additionally, the waveguide structure should provide a good carrier transfer to the quantum wells. The quantum wells should present high radiative efficiency and spectrally narrow emission. Finally, the SCH ensemble should provide high optical gain. Being able to address properly such issues should help to improve the performance of the next generation of AlGaN lasers, which are still outperformed by InGaN lasers in terms of efficiency.

In this paper, we analyze optical properties of AlGaN-based SCHs designed for the implementation of electron beam pumped UV lasers. We study their photoluminescence under various injection levels and compare the internal quantum efficiency (IQE) of the spontaneous emission and the lasing threshold power density, to determine the limiting factors. We discuss the effect of carrier localization and carrier diffusion on the device performance.

The samples were grown by plasma-assisted molecular beam epitaxy on bulk GaN substrates.[7] The three samples under study consist of a 10-period of GaN/Al$_{0.1}$Ga$_{0.9}$N multi-quantum well (MQW) inserted in an Al$_{0.1}$Ga$_{0.9}$N/Al$_{0.2}$Ga$_{0.8}$N waveguide to form an SCH. A schematic description is presented in Fig. 1(a), and the dimensions of the layers are listed in Table 1. Sample S1 presents chemically sharp heterointerfaces (Al mole fraction profile in Fig.

1(b)), whereas the $Al_{0.1}Ga_{0.9}N/Al_{0.2}Ga_{0.8}N$ heterointerfaces in sample S2 are linearly graded along 35.2 nm (profile in Fig. 1(c)) to implement a graded index separate confinement heterostructure (GRINSCH). Finally, the sample S3 consists of an asymmetric GRINSCH, where the top cladding layer has higher Al content and the top graded region is larger, reaching the top of the MQW (profile in Fig. 1(d)). The GRINSCH design aims to decrease the bandgap bending due to spontaneous and piezoelectric polarisation at the $Al_{0.1}Ga_{0.9}N/Al_{0.2}Ga_{0.8}N$ interface in order to promote the diffusion of carriers to the MQW. The asymmetric GRINSCH was implemented to force the carriers generated in the top cladding layers to diffuse to the MQW. A more detailed explanation of the designs with simulations of the band diagrams were reported elsewhere.[7] In particular, we studied the confinement of the optical mode in the waveguide using a commercial finite-element analysis software (Comsol Multiphysics). The refractive indices used as input parameters were extracted from refs. [8,9]. The resulting mode distribution profiles are displayed in Figs. 1(b-d), superimposed to the Al content profiles. The optical mode profiles are very similar for S1 and S2, whereas the mode is slightly shifted towards the substrate in S3 due to the higher Al content in the top cladding layer. If we consider the 10 quantum wells as active elements, the optical confinement factors are 3.84%, 3.73%, and 3.79% for S1, S2, and S3, respectively. If the integration is extended along the MQW (including the barriers), the optical confinement factors are 36.3%, 36.5%, and 35.3%, respectively. These values, summarized in Table 1, confirm the optical similarity of the three samples.

Photoluminescence (PL) spectroscopy under low injection conditions was measured by excitation with a continuous-wave frequency-doubled solid-state laser ($\lambda = 244$ nm), with an optical power of 10 μW focused on a spot with a diameter of ≈ 100 μm. PL measurements under pulsed excitation used a Nd-YAG laser (266 nm, 0.5 ns pulses, repetition rate of 8 kHz). In all the cases, the luminescence was collected by a Jobin Yvon HR460 monochromator equipped with a UV-enhanced charge-coupled device (CCD) camera.

For lasing threshold measurements, the samples were mechanically cleaved along *m* planes of the GaN substrate, to form laser bars with a resonator length of 1 mm. Excitation was provided by the Nd-YAG laser. A cylindrical lens was inserted in the optical path to shape the laser beam into a 100-µm-wide stripe perpendicular to the sample facets. In this configuration, the lateral confinement of light is obtained by gain guiding, i.e. the population-inverted region is determined by the width of the pumping laser stripe, which generates a gradient of gain and refractive index. The emission was collected from the edge of the laser bar. This characterization is described in detail in ref. [7]. The values of lasing threshold obtained from such measurements are listed in Table 1.

The performance of a UV emitter can be described by its external quantum efficiency (EQE), which is the ratio between the number of photons detected and the number of photons (optical pumping) or electrons (electrical pumping) injected in the device. In an LED, the EQE can be expressed as the product of the injection efficiency ($\eta_{inj}$, ratio of carriers reaching the light-emitting region), the radiation efficiency ($\eta_{rad}$, radiative recombination ratio) and the light extraction efficiency ($\eta_{ext}$, ratio of photons coming out of the sample). Note that the above-defined $\eta_{inj}$ is different than the injection efficiency generally used in laser diodes, which describes the fraction of current above the threshold that results in radiative recombination. Also in the case of a laser, a relevant parameter is the differential EQE, $\Delta_{EQE}$, extracted from the slope of the output power vs. pumping power curve above the lasing threshold. However, if we want to assess the material properties, it is more interesting to refer to the internal quantum efficiency, which is defined as IQE = $\eta_{inj} \times \eta_{rad}$. Unlike the EQE and $\Delta_{EQE}$, the IQE is independent of the geometrical properties of the laser, such as the cavity length or the ridge width. Hence, it is a good parameter to compare the material quality. The IQE is generally extracted from temperature-dependent PL measurements. It is often assumed that the IQE at a certain temperature (T) can be calculated as

$$\text{IQE}(T) = \frac{I_{PL}(T)}{I_{PL}(T = 0K)} \quad (1)$$

where $I_{PL}(T)$ is the integrated PL intensity at T. This estimation is based on the hypothesis that the PL intensity saturates at low temperature, since carrier freeze out prevents them from reaching non-radiative recombination centers.

To obtain the intrinsic IQE, associated with the material properties, the excitation should introduce as little perturbation as possible. Therefore, we have performed low-injection temperature-dependent PL measurements, with the results presented in Figs. 2(a-c) for S1, S2 and S3, respectively. To understand the spectra, we must keep in mind the penetration depth of the pumping laser: In view of the absorption coefficient of GaN, 80% of the absorption occurs in the TOC+TIC layers, and carriers have to diffuse towards the MQWs. Therefore, the observed transitions can be assigned to recombination in the TOC and TIC layers, in the MQW, and to a low energy band due to donor-acceptor recombination (DAP), as indicated in the figures. In the three samples under study, the ratio of integrated PL intensity from the MQW at room temperature and at low temperature is systematically lower than 1%. Such low values are generally observed in GaN-based MQWs measured under low injection conditions.[10,11]

Looking at the PL line assigned to the MQW in Fig. 2, the spectral evolution of the PL peak presents an S shape, outlined with dashed lines in the figures. This behavior is assigned to the localization of carriers in potential fluctuations in the QWs,[12] which can be due to fluctuations of the QW thickness or alloy inhomogeneities in the QW barriers. Figure 2(d) shows the evolution of the MQW PL peak energy as a function of temperature for the three samples, and the high temperature trend was fitted with Varshni's equation.[13] The deviation from this trend observed at low temperature provides an estimation of the carrier localization energy, which varies from $E_{loc}$ = 40±2 meV for S1 to 20±1 meV for S2 and 17±1 meV for S3.

A similar study applied to the TIC line, assigned to recombination in the $Al_{0.1}Ga_{0.9}N$ inner cladding, leads to $E_{loc}$ = 42±4 meV for S1 to 37±4 meV for S2 and 19±3 meV for S3. This points to the fact that carrier localization in the MQW is induced by alloy inhomogeneities in the barriers rather than thickness fluctuations of the wells. We note also that the localization energy in the TOC layers, with higher Al content, increases to the range of 44-53 meV in all the samples, which hinders the diffusion of carriers towards the MQW at low temperature and explains the presence of the TOC-related lines in all the spectra in Figs. 2(a-c).

It is known that the IQE of III-nitrides is strongly dependent on the injection intensity.[14–16] Under operating conditions, the carrier density in the structure is high enough to saturate non-radiative recombination paths and screen potential barriers associated with defects or interfaces. Emulating such conditions requires high pumping densities. Therefore, we have studied the IQE of the SCH as a function of temperature and excitation power density using the method described by Yamada et al.[17]:

$$IQE_{SCH}(T) = \frac{\eta_{PL}(T)}{max[\eta_{PL}(T=0K)]} \qquad (2)$$

where $\eta_{PL}$ is the PL efficiency, defined as the integrated PL intensity divided by the excitation power. The PL spectra were measured for various excitation powers, as presented for S1 at 6 K and 300 K in Figs. 3(a) and (b). The narrow lines labeled "*" that appear at high excitation power are assigned to stimulated emission. Such a phenomenon is not expected in a sample without cleaved facets, but it occurs due to the feedback provided by reflection at cracks,[18] which appear due to the strong lattice mismatch between different layers.

The calculation of $IQE_{SCH}$ considers the integrated intensity of the whole PL spectrum. However, the contribution of recombination in the cladding layers to the total intensity can be important, particularly at low temperature, due to poor carrier diffusion to the MQW. In view of lasing, radiative recombination in the cladding layers should be considered as losses.

Therefore, we have calculated IQE$_{MQW}$ integrating only the optical signal from the MQW, which can be extracted using a Lorentzian fit to remove the other contributions, with the maximum PL efficiency still given by the integrated PL over the whole spectrum at low temperature, i.e.

$$\text{IQE}_{MQW}(T) = \frac{\eta_{PL\ MQW}(T)}{\max[\eta_{PL}(T=0K)]} \qquad (3)$$

where $\eta_{PL\ MQW}$ is the integrated PL intensity of the MQW divided by the pumping power and $\eta_{PL}$ is the total integrated PL intensity divided by the pumping power. Comparing IQE$_{SCH}$ and IQE$_{MQW}$, we can extract information about the efficiency of the carrier transfer from the waveguide to the MQW. Figure 3(c) displays the results of IQE$_{SCH}$ and IQE$_{MQW}$ for S1. There is a significant deviation at low temperature, but the two values are very close at room temperature due to the thermally enhanced carrier diffusion towards the MQW.

Figures 4(a-c) show IQE$_{MQW}$ as a function of pump power and temperature for the three architectures under study. At 6 K, the highest IQE$_{MQW}$ is about 49%, 56% and 85%, for S1, S2 and S3, respectively (data summarized in Table 1). Note that the highest IQE$_{SCH}$ at low temperature is considered to be 100%, so that the highest IQE$_{MQW}$ gives direct information on the efficiency of the carrier transfer to the MQW. The difference between samples can be explained by their band diagram[7]: The GRINSCH in S2 and S3 promotes the diffusion of carriers along the growth axis towards the MQW, which is one of the factors that limit the IQE$_{MQW}$ at cryogenic temperatures. However, at 300 K, the IQE$_{MQW}$ is about 22%, 11% and 5% for S1, S2 and S3, respectively, i.e. the trend is reversed compared to that at 6 K. This suggests that the higher carrier mobility at room temperature promotes not only carrier collection in the MQW, due to the enhanced diffusion length along the growth axis, but also their trapping in nonradiative centers, due to the enhanced in-plane diffusion length. As the acceleration of non-radiative processes is higher in samples with lower carrier localization, the performance of S3 is penalized with respect to S2 and S1.

To get a deeper insight into this issue, we have studied the variation of the maximum $IQE_{MQW}$ as a function of temperature in the three samples, with the result illustrated in Fig. 4(d). The $IQE_{MQW}$ remains constant at low temperature, and drops sharply for higher temperature due to the activation of non-radiative recombination paths. It is interesting to note that the most severe thermal quenching is observed in sample S3, which features the highest $IQE_{MQW}$ at 6 K (see Fig. 2).

The dashed lines in Fig. 4(d) are fits assuming the dominance of a monoexponential non-radiative process, so that the variation with temperature is described by[19]

$$IQE(T) = \frac{IQE(T=0)}{1 + A\exp(-E_a/kT)} \tag{4}$$

where $E_a$ is the activation energy of the non-radiative process, A is a fitting parameter that is determined by the ratio between the radiative and the non-radiative recombination times,[19] and kT is the thermal energy. In the figure, $E_a$ = 22±5 meV, 16±9 meV, and 14±2 meV for samples S1, S2 and S3, respectively. Even if the energy values are smaller than the carrier localization energy extracted in Fig. 2(d), which can be explained by the different excitation conditions, the trend obtained here is consistent with the trend of the carrier localization energy extracted from the variation of the PL peak energy with temperature. On the other hand, the values of A extracted from the fits are A = 2.4±0.6, 6±3, and 12±3 for samples S1, S2 and S3, respectively. Even if the error bars are large, the trend points to an acceleration of non-radiative phenomena in sample S2 and particularly in sample S3 with respect to S1, which might indicate a higher density of point defects behaving as non-radiative recombination centers. These results suggest that the enhanced mobility of carriers at room temperature in S2 and S3 promotes not only carrier collection in MQWs but also their reaching non-radiative traps, which are more effective due to the reduced localization in the MQW. The particularly high value of A in S3 might indicate a higher density of point defects in the structure, which could be explained by the higher aluminum content in the upper layers.

Comparing the values of $IQE_{MQW}$ and lasing threshold at room temperature, summarized in Table 1, we observe that these two parameters are not correlated. Higher $IQE_{MQW}$ at room temperature seems to be associated with higher carrier localization in the MQW, which is explained by the fact that potential fluctuations prevent carriers from reaching non-radiative centers. A decorrelation of the lasing threshold and the $IQE_{MQW}$ is expected in samples presenting potential fluctuations in the wells, since they lead to inhomogeneous broadening of the gain spectrum.[20] In that case, higher pumping is required to attain a high-enough gain to overcome the losses, consequently increasing the lasing threshold, even if the MQW is more radiative efficient. Inhomogeneous spectral broadening of the gain generally manifests as inhomogenous broadening the photoluminescence. However, this is not the case here. If we compare the full width at half maximum (FWHM) of the MQW PL peaks in Figs. 2(a-c) with the lasing power density threshold, displayed in Table 1, the values do not follow the same tendency (e.g. S2 has a narrower emission line than S3 in spite of having the same lasing threshold).

The lasing threshold depends on more factors, namely the attainment of a carrier concentration high enough to ensure population inversion and a gain level that compensates the optical losses. In the three structures under study, the waveguide was designed to present similar optical confinement. The laser bars were cleaved in the same manner and the cavity length was the same, long enough to ensure that the distributed mirror losses are relatively small. Therefore, the variation of the threshold should not be associated with the optical properties of the waveguide, but rather to the difference in carrier injection and material quality.

To gain further insight into the role of the carrier mobility and localization on the device performance, Fig. 4(e) displays the variation of the lasing threshold as a function of temperature in the three samples. The decrease of the threshold when reducing the temperature is similar in samples S1 and S2, where an exponential fit (expected empirical dependence[21]) reveals an activation energy $E_{th}$ = 12-13 meV, whereas it increases to $E_{th}$ = 38 meV for S3. To understand the high threshold of S3 at low temperature, we must remind most of the absorption occurs in the TOC+TIC layers. Therefore, the higher losses that appear when cooling down can relate to the alloy fluctuations and higher density of non-radiative point defects in the Al-rich topmost AlGaN layers. In contrast, at high temperature, carriers have enough energy to follow the potential ramp created by the GRINSCH and diffuse efficiently to the MQW.

In summary, this work discusses the different trends between internal quantum efficiency and power density lasing threshold in AlGaN-based separate confinement heterostructures. To assess the performance of these structures, it is important to analyze both the radiative efficiency and the carrier injection. At room temperature, carrier localization in the quantum wells leads to an enhancement of the radiative efficiency. However, our results show that the lasing threshold is more sensitive to the injection efficiency than to the radiative efficiency. Therefore, designs including graded alloys result in a reduction of the lasing threshold, in spite of their lower internal quantum efficiency.

## ACKNOWLEDGEMENTS

This work is supported by the French National Research Agency via the UVLASE program (ANR-18-CE24-0014), and by the Auvergne-Rhône-Alpes region (PEAPLE grant).

## DATA AVAILABILITY STATEMENT

The data that support the findings of this study are available from the corresponding authors upon reasonable request.

# TABLE

**Table 1.** Description of samples under study: thickness and Al content of the layers, following the general design in Fig. 1(a). Optical confinement factor integrating the optical mode in the wells (OCF$_W$) and along the whole MQW structure (OCF$_{MQW}$). Lasing power density threshold at room temperature (RT). Maximum IQE$_{MQW}$ at 6 K and room temperature and maximum IQE$_{SCH}$ at room temperature (note that the maximum IQE$_{SCH}$ at 6 K is considered to be 100%). Lasing threshold measured in mechanically cleaved 1-mm long cavities when optically pumped with a Nd-YAG laser.

|  | S1 | S2 | S3 |
| --- | --- | --- | --- |
| TOC | 44.4 nm Al$_{0.2}$Ga$_{0.8}$N | 26.4 nm Al$_{0.2}$Ga$_{0.8}$N | 44.2 nm Al$_{0.3}$Ga$_{0.7}$N |
| Graded | -- | 35.2 nm | 48.7 nm |
| TIC | 59.5 nm Al$_{0.1}$Ga$_{0.9}$N | 41.4 nm Al$_{0.1}$Ga$_{0.9}$N | -- |
| MQW | 10×(1.4 nm GaN /9.8 nm Al$_{0.1}$Ga$_{0.9}$N) | 10×(1.3 nm GaN /9.7 nm Al$_{0.1}$Ga$_{0.9}$N) | 10×(1.3 nm GaN /9.7 nm Al$_{0.1}$Ga$_{0.9}$N) |
| BIC | 35.5 nm Al$_{0.1}$Ga$_{0.9}$N | 17.6 nm Al$_{0.1}$Ga$_{0.9}$N | 17.7 nm Al$_{0.1}$Ga$_{0.9}$N |
| Graded | -- | 35.2 nm | 35.4 nm |
| BOC | 355 nm Al$_{0.2}$Ga$_{0.8}$N | 332 nm Al$_{0.2}$Ga$_{0.8}$N | 334 nm Al$_{0.2}$Ga$_{0.8}$N |
| OCF$_W$ (%) | 3.84 | 3.73 | 3.79 |
| OCF$_{MQW}$ (%) | 36.3 | 36.5 | 35.3 |
| RT Lasing threshold (kW/cm$^2$) | 210 | 180 | 180 |
| max[IQE$_{MQW}$(T = 6 K)] (%) | 49 | 56 | 86 |
| max[IQE$_{MQW}$(T = 300 K)] (%) | 22 | 11 | 5.0 |
| max[IQE$_{SCH}$(T = 300 K)] (%) | 26 | 13 | 5.5 |

**FIGURE CAPTIONS**

**Figure 1.** (a) Scheme of the samples. Bottom to top: Bulk GaN substrate, bottom outer cladding (BOC), graded layer, bottom inner cladding (BIC), MQW, top inner cladding (TIC), graded layer, top outer cladding (TOP). Alloy concentration and simulation of the optical mode profile along the growth direction in samples (b) S1, (c) S2 and (d) S3.

**Figure 2.** PL spectra recorded at low injection conditions for samples (a) S1, (b) S2 and (c) S3. The emission assigned to the outer claddings (OC), inner claddings (IC), MQW, and donor-acceptor pair (DAP) recombination are indicated in the figures. Solid lines associate the MQW line with its first and second phonon replicas. Dashed curves describe the evolution of the MQW emission with temperature. The narrow line labeled "*" that appears at low temperature in (a) is assigned to stimulated emission. (d) Variation of the MQW peak energy as a function of temperature, compared with the trend given by Varshni equation (dashed lines) to estimate the carrier localization energy, $E_{loc}$.

**Figure 3.** For sample S1, variation of the PL spectrum as a function of excitation power, measured at (a) 6 K and (b) 300 K. The narrow line labeled "*" that appears at high pumping powers is assigned to stimulated emission. (c) Estimation of the $IQE_{SCH}$ and $IQE_{MQW}$ as a function of the excitation power density, at 6 K and 300K.

**Figure 4.** $IQE_{MQW}$ as a function of the excitation power density at different temperatures, calculated with method 3 for (a) S1, (b) S2, (c) S3. (d) Maximum $IQE_{MQW}$ as a function of the inverse temperature for S1, S2 and S3. (e) Variation of the lasing threshold as a function of temperature.

**Figure 1**

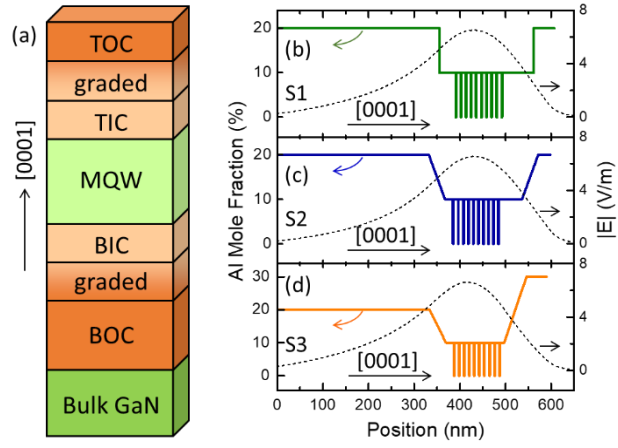

**Figure 2**

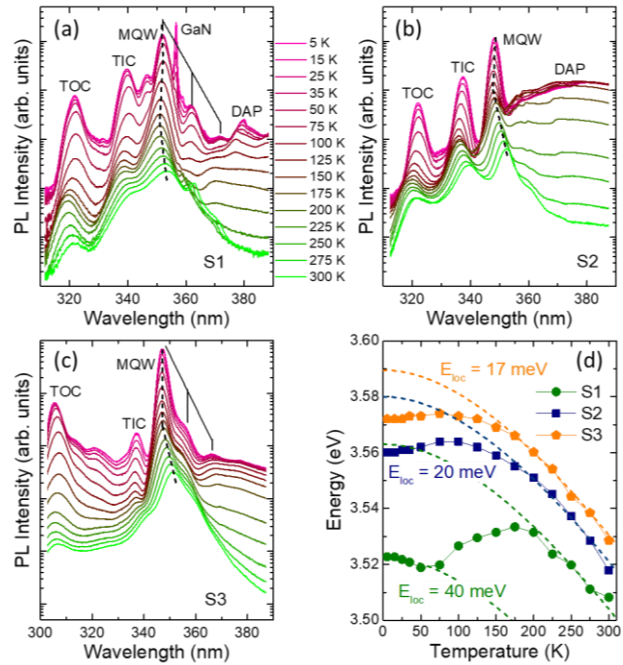

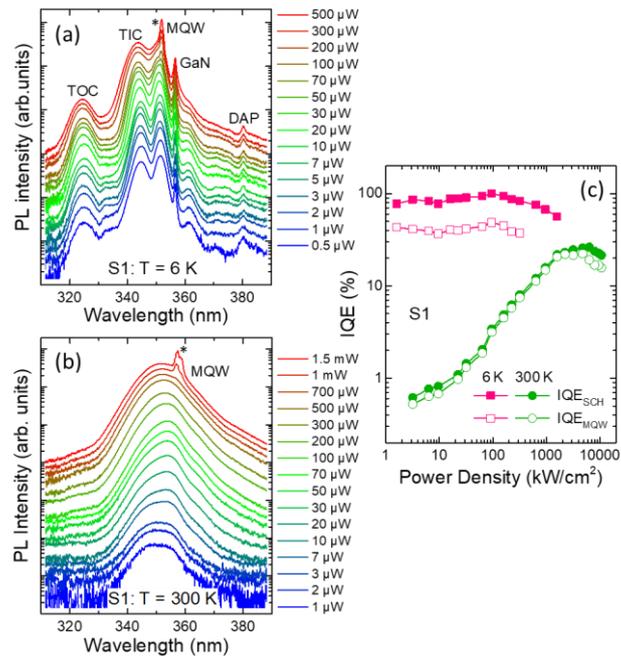

**Figure 4**

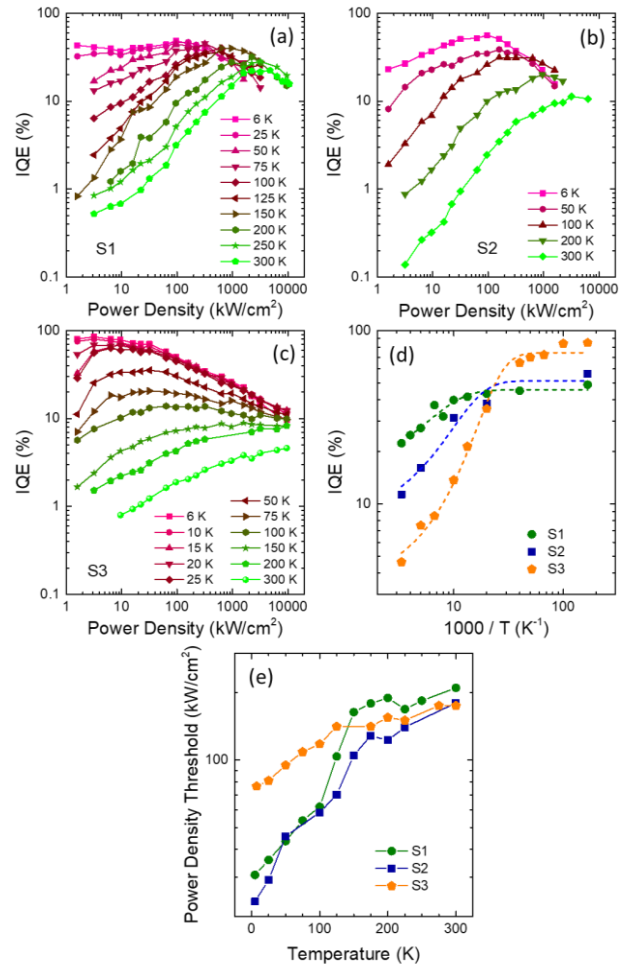